\documentclass[aps,twocolumn,pra,superscriptaddress]{revtex4}
\pdfoutput=1
\usepackage[utf8]{inputenc}
\usepackage{epsfig,graphicx,times}
\usepackage{amstext}

\usepackage{amsmath}
\usepackage{amssymb}
\usepackage{graphicx}
\usepackage{latexsym}
\usepackage{bm}
\usepackage[colorlinks,citecolor=blue, linkcolor=blue,hyperindex,CJKbookmarks,pdftex]{hyperref}
\usepackage[toc,page,title,titletoc,header]{appendix}
\usepackage{epstopdf}

\begin{document}
\title{Giant atom induced  zero modes and localization in the nonreciprocal Su-Schrieffer-Heeger chain}
\author{Junjie Wang}
\affiliation{Center for Quantum Sciences and School of Physics, Northeast Normal University, Changchun 130024, China}
\author{Fude Li}
\affiliation{Center for Quantum Sciences and School of Physics, Northeast Normal University, Changchun 130024, China}
\author{X. X. Yi}
\email{yixx@nenu.edu.cn}
\affiliation{Center for Quantum Sciences and School of Physics, Northeast Normal University, Changchun 130024, China}
\affiliation{Center for Advanced Optoelectronic Functional Materials Research, and Key Laboratory for UV-Emitting Materials and Technology of
Ministry of Education, Northeast Normal University, Changchun 130024, China}

\date{\today}

\begin{abstract}
A notable feature of non-Hermitian systems with skin effects is the sensitivity of their spectra  and eigenstates to the boundary conditions. In the literature, three types of boundary conditions--periodic boundary condition, open boundary condition and a defect in the system as a boundary,  are explored. In this work we introduce the other type of boundary condition provided by a giant atom.  The giant atom couples to a nonreciprocal Su-Schrieffer-Heeger chain at two points and plays the role of defects. We study the spectrum and localization of eigenstates of the system and find that the giant atom can induce asymmetric zero modes. A remarkable feature is that bulk states might localize at the left or the right chain-atom coupling sites in weak localization regimes. This bipolar localization leads to Bloch-like states, even though translational invariance is broken. Moreover, we find that the localization is obviously weaker than the case with two small atoms or  open boundary conditions even in strong coupling regimes. These intriguing results indicate that nonlocal coupling of the giant atom to a nonreciprocal SSH chain  weakens localization of the eigenstates. We also show that the Lyapunov exponent in the long-time dynamics in real space can act as a witness  of the localized bulk states.
\end{abstract}


\maketitle
\section{Introduction}\label{sec1}
In closed quantum systems, observables like Hamiltonians are  represented by Hermitian operators. This is not the case for real systems that are  inevitably  coupled to  environments.  It is widely accepted that  open quantum systems can be described effectively by  non-Hermitian operators on the basis of  the quantum trajectory approach. Recently non-Hermitian systems have attracted a lot of attention~\cite{Majorana1931,Feshbach1958,Nelson1996,Boettcher1998,Berry2004,Moiseyev2011,Brody2013} especially in the field of topological physics. Interesting  features are predicted and observed such as the topological insulator laser~\cite{Rechtsman2018, Szameit2018, Khajavikhan2018}, new topological invariants in non-Hermitian systems~\cite{Liang2013,Lee2016,Leykam2017,Z. Gong2018,Nori2019,Chen2018,Lieua2018,Shen2018}, and  the breakdown of the conventional bulk-boundary correspondence~\cite{Xiong2018,Bergholtz2018,Yaoa2018,Yaob2018,Wang2019a,Wang2019b,JiangpingHu2020,
ChenFang2020,Zhesen2020,Slager2020,Masatoshi2020,Torres2018,Thomale2019,Jiangbin2020a,Jiangbin2020b,Jiangbin2020c,Narayan2022, ShuangZhang2021,Neupert2020,Xue2020,Thomale2020,Coulais2019}.

The systems with non-Hermitian skin effect (NHSE) are very sensitive with respect to boundary conditions. For example, the properties of both spectrum and eigenstates of non-Hermitian systems may change dramatically   by changing the boundary conditions from periodic to open ones~\cite{Yaoa2018}. In between, a defect introduced into the system could also play the role of boundary~\cite{Jiangbin2020,Murakami2021,Chen2021,Chen2020,Roccati2021}. Non-Hermitian impurity physics~\cite{Jiangbin2020} was developed theoretically in
a non-reciprocal lattice, where the authors found  a new type of  steady-state localization  characterized by scale-free accumulation of
eigenstates with localization lengths   proportional to the system size.
The scale-free localization can be interpreted by the generalized non-Bloch band theory~\cite{Murakami2021,Chen2021}, and strong defect-system couplings  might induce an effective boundary in a periodic system with properties very similar to systems with open boundary condition (OBC)~\cite{Chen2020,Roccati2021,Shen2011,Balents2015}.

Giant atom was first physically realized by nonlocally coupling a transmon qubit to surface acoustic wave via interdigitated transducer~\cite{Delsing2014}. Because of the slow velocity of the acoustic wave, the size of the transmon is biger than wavelength, so the qubit is called  giant atom. On the other hand, giant atom such as superconducting qubit has played an important role in superconducting quantum circuits. It  can be nonlocally coupled to a waveguide at multiple points~\cite{Ciccarello2022,Oliver2020,Leek2017,Lehnert2020,XinWang2022,Nori2021,Liu2020,Bachelard2022,Delsing2020} and act as an effective boundary for the waveguide. The giant atom might induce  chiral bound states for the Hermitian systems of waveguide~\cite{Nori2021,Liu2020}. These begs the question that what properties of the system will change if a giant atom couples to a non-Hermitian topological chain and what new physical phenomena emerge. Furthermore, it is interesting to ask if there is a difference between the localisation
induced by a giant atom or other impurities and defects.

In this work, we will answer these questions and focus on spectrum structures and the localization of eigenstates in a system composed of a giant atom and a nonreciprocal Su-Schrieffer-Heeger(SSH) chain. We find that the giant atom can induce asymmetric zero modes. Bulk states may localize at the left or the right chain-atom coupling sites, so we call this ``a bipolar localization".
Further examination shows that strong atom-chain coupling  can not induce transitions from skin-free  to skin states in the bulk states. This suggests that the localization  is weaker than that with OBC even in strong coupling regimes, and  nonlocal couplings weaken localization  of the eigenstates. We also show that the feature of localization  of the bulk states can be captured by  the  dynamics in real space.

The paper is organized as follows. In Sec. {\rm\ref{sec2}}, we introduce a model to describe the system composed of a giant atom and a nonreciprocal SSH chain and calculate the spectrum of the system. In Sec. {\rm\ref{sec3}}, we derive an analytical
expression for the zero modes induced by the giant atom. Two cases, one with two small atoms and the other with a giant atom, are compared and discussed in details.  In Sec. {\rm\ref{sec4}}, we mainly focus on  the localization feature of the bulk eigenstates. In Sec. {\rm\ref{sec5}}, we explore the relation  between the Lyapunov exponent and the localization of the bulk states. Finally, we summarize our results in Sec. {\rm\ref{sec6}}.

\section{model and spectrum}\label{sec2}
\begin{figure}[tbp]
\centering
\includegraphics[width=9cm]{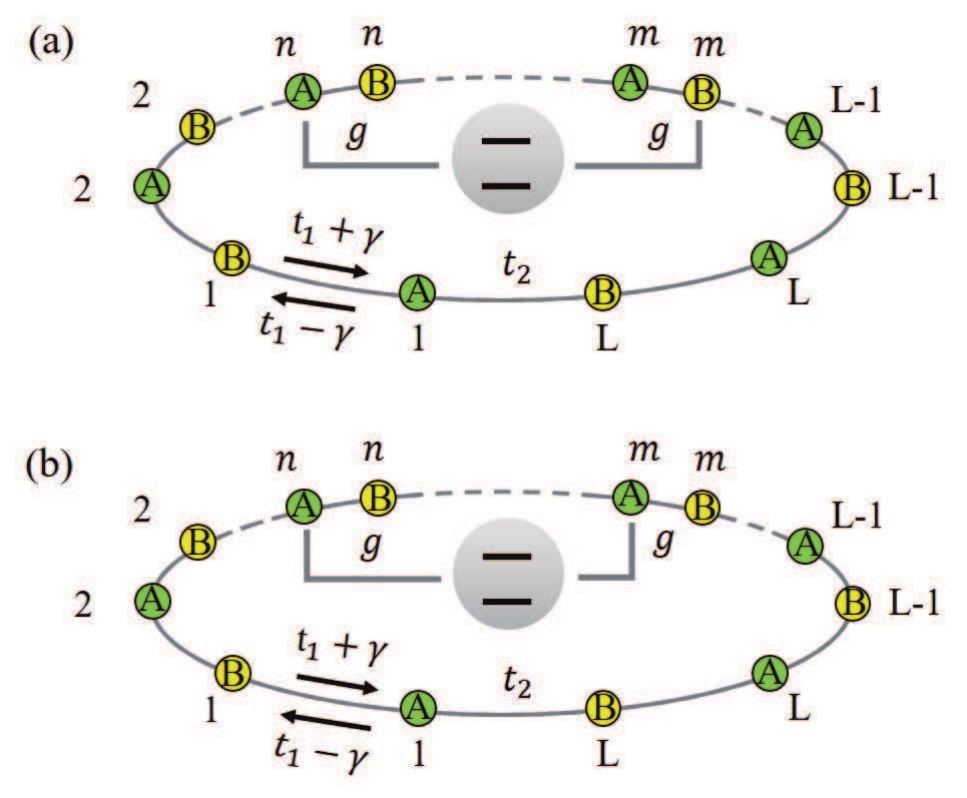}
\caption{Sketch of the system. (a) A giant atom coupled to a nonreciprocal SSH chain via A-B coupling. (b) A giant atom coupled to a nonreciprocal SSH chain via A-A coupling.}
\label{f1}
\end{figure}
As depicted in Figs.~\ref{f1} (a) and (b), we consider  a nonreciprocal SSH chain with periodic boundary
conditions (PBC) in real space  described by
\begin{equation}
\begin{aligned}
H_{\mathrm{SSH}}=&\sum_{l=1}^L[(t_1+\gamma)\hat C^{\dag}_{A,l}\hat C{_{B,l}}+(t_1-\gamma)\hat C{^{\dag}_{B,l}}\hat C{_{A,l}}\\
&+t_2\hat C{^{\dag}_{A,l+1}}\hat C{_{B,l}}+t_2\hat C{^{\dag}_{B,l}}\hat C{_{A,l+1}}],
\label{1}
\end{aligned}
\end{equation}
where the chain consists of $L$ unit cells, and the staggered nearest-neighbor hopping amplitudes are $t_1\pm\gamma$ and $t_2$, respectively. The asymmetry of hopping amplitudes $(\gamma\neq 0)$ leads to the non-Hermiticity of
the system. $\hat C^{\dag}_{A(B),l}$ and $\hat C_{A(B),l}$ are the creation and annihilation operators for the sublattice site $A(B)$ at site $l$. To have  an effective boundary, we introduce  a two level giant atom coupling at two points to a nonreciprocal SSH chain via $A-B$ (or $A-A$) couplings. The interaction  Hamiltonian for such a system can be  written as
\begin{equation}
\begin{aligned}
H_{I, AB} &=g|e\rangle\langle g|(\hat C_{A, n}+\hat C_{B, m})+\mathrm{ H.c. }, \\
H_{I,AA}&=g|e\rangle\langle g|(\hat C_{A,n}+\hat C_{A,m})+\mathrm{H.c.},
\label{2}
\end{aligned}
\end{equation}
where $g$ denotes the atom-chain coupling strength. $|g\rangle$ and $|e\rangle$ are the ground state and the excited state of the giant atom, respectively. The Hamiltonians of atom-chain coupling read
\begin{subequations}
\begin{equation}
H_{AB}=H_{\mathrm{SSH}}+H_{I,AB},
\label{3a}
\end{equation}
\begin{equation}
H_{AA}=H_{\mathrm{SSH}}+H_{I,AA}.
\label{3b}
\end{equation}
\end{subequations}
In order to derive a condition for the emergence of the zero mode, in the single-excitation subspace, the eigenstate of the Bloch Hamiltonian in momentum space can be expressed as
\begin{equation}
\begin{aligned}
|\psi \rangle  = {U_e}|e,G\rangle  + \sum\limits_k A_k{\hat C_{A,k}^\dag } |g,G\rangle  + \sum\limits_k {{B_k}} \hat C_{B,k}^\dag |g,G\rangle,
\label{4}
\end{aligned}
\end{equation}
where $\hat C^{\dag}_{A(B),k}$ and $\hat C_{A(B),k}$ are the creation and annihilation operators for the sublattice site $A(B)$ at site $k$. $|G\rangle$ denotes the ground state of the SSH chain. Making use of the time-independent Schr{\"o}dinger equation $H(k)|\psi\rangle=E|\psi\rangle$, where $H(k)$ is the Hamiltonian of the system in the momentum space (See Appendix~\ref{A1} for detail), we obtain the eigenvalues of the system with  $A-B$ couplings
\begin{equation}
\begin{aligned}
E&=\frac{2 g^{2}}{L} \sum_{k} (E+t_{1} \cos [k(m-n)]-\gamma \sin [k(m-n)]\\
&+t_{2} \cos [k(m-n+1)])/(E^{2}-\omega_{k}^{2}) \\
\label{5}
\end{aligned}
\end{equation}
with
\begin{equation}
\omega_k=\sqrt{(t_1+\gamma +t_2 e^{-ik})(t_1-\gamma +t_2 e^{ik})}.
\label{6}
\end{equation}
\begin{figure}[tbp]
\centering
\includegraphics[width=9cm]{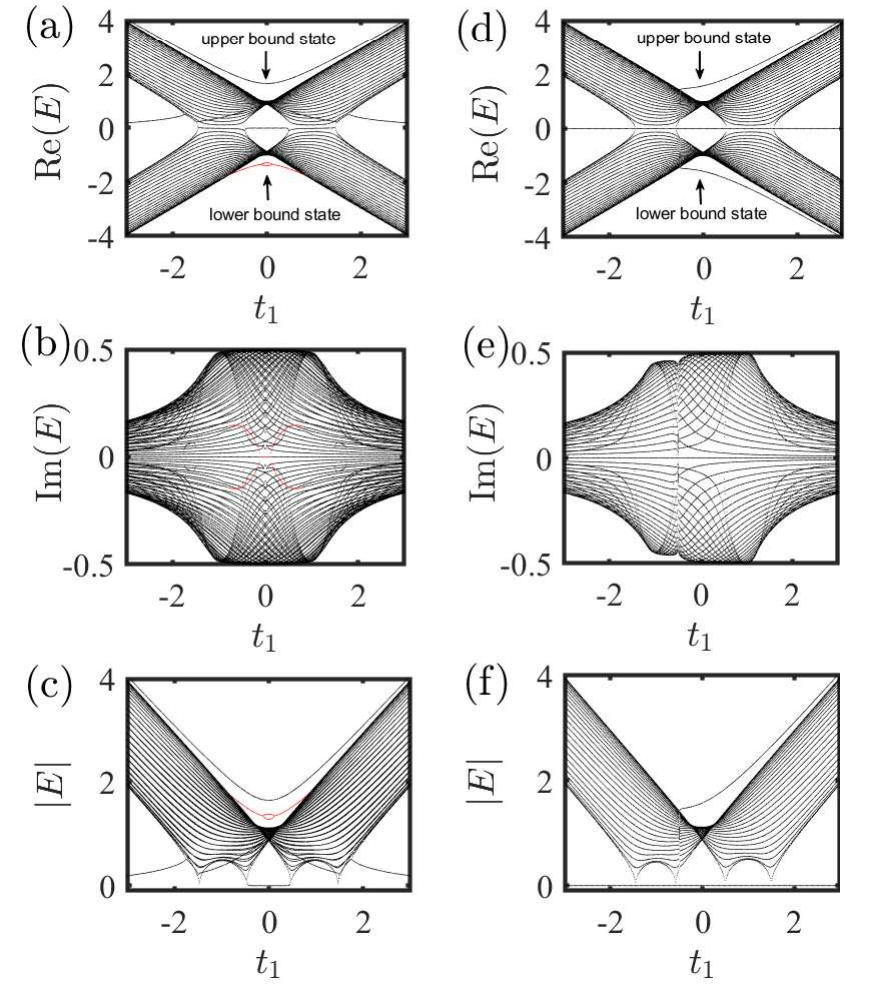}
\caption{Spectrum as a function of $t_1$ in (a,b,c) with $A-B$ and (d,e,f) with $A-A$ coupling, respectively. First (second) row show real (imaginary) part of the spectrum, while last row shows absolute value of the spectrum. The results are obtained by numerically solve the Sch\"odinger equation. The chosen parameters are $L = 50, m=26, n=25, g=1, t_1=0.2, t_2 = 1,$ and $\gamma = 0.5$.}
\label{f2}
\end{figure}
Setting $E = 0$, we can derive a condition for the emergence of the zero mode, i.e.,  $t_1\in[-t_2+\gamma, t_2-\gamma]$. In the absence of giant atoms, the different topological phases for nonreciprocal SSH model can be distinguished by the
winding number~\cite{Chen2018} $v=\frac{1}{\pi} \int_{-\pi}^{\pi} d k\left\langle\phi^{L}\left|i \partial_{k}\right| \phi^{R}\right\rangle$. It satisfies $v=1$ for $t_1\in[-t_2+\gamma, t_2-\gamma]$. Note that the condition for the emergence of the zero mode is identical to the non-trivial phase boundary of systems without giant atoms. (See Appendix~\ref{A1} for analytical results and Fig.~\ref{f2}\,(a-c) for numerical simulations). Except for the modes in the gap of the real part of the energy, there are other eigenvalues outside (lying  above and below) the continuous bands as shown in Fig.~\ref{f2}\,(a). We call the corresponding eigenstates upper and lower bound states, respectively. Interestingly, eigenvalues form a close circle (red loop) in Fig.~\ref{f2}\,(c), because the system crosses two exception points as the increase of parameter $t_1$. Combined with Fig.~\ref{f2}\,(a,b), it is clearly shown that complex conjugate pair of eigenvalues become purely real when crossing the left exception point, and adjacent real eigenvalues merge into complex conjugate pairs when crossing the right exception point.

For A-A coupling, the eigenvalue satisfies
\begin{equation}
E=\frac{2 g^{2}}{L} \sum_{k}\left(\frac{E(1+\cos [k(m-n)])}{E^{2}-\omega_{k}^{2}}\right).
\label{7}
\end{equation}
Obviously, $E = 0$ is always the solution of Eq.~\eqref{7}, which can be seen in the numerical spectra of Hamiltonian $H_{AA}$ (Fig.~\ref{f2}\,(d-f)). In Fig.~\ref{f2}\,(d), we find that the upper and lower eigenvalues merge into the continual band as $t_1$  decreases, which indicates that the corresponding bound states vanished.

\section{Zero mode and bound states}\label{sec3}

\begin{figure}[tbp]
\centering
\includegraphics[width=8.8cm]{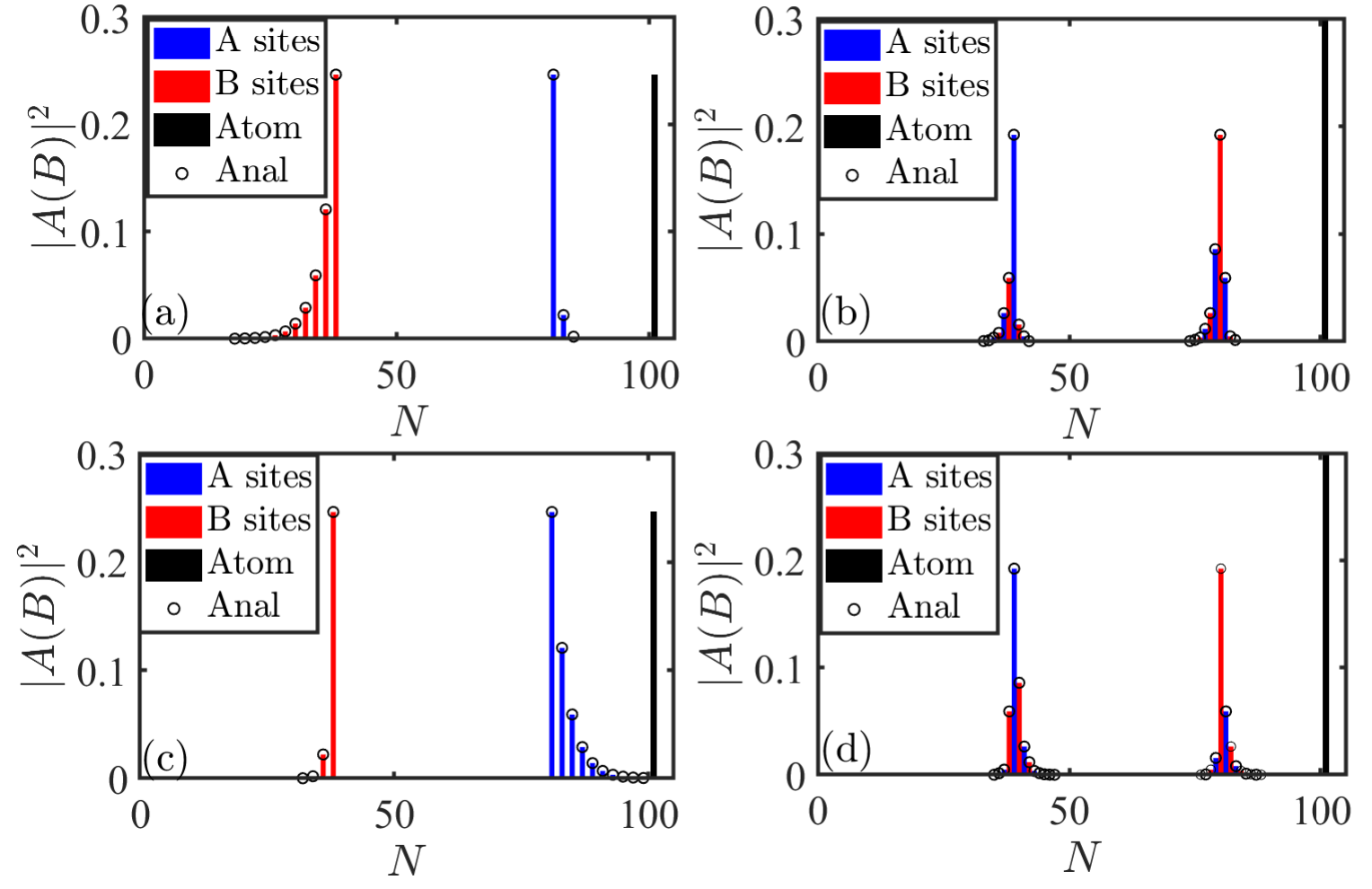}
\caption{The probability distributions of bound states for the system~\eqref{3a} with $A-B$ coupling as a function of lattice site $N$. (a) and (c) are plotted for the probability distribution of zero mode. (b) and (d) are plotted for the probability distribution of the upper bound state. Here, $t_1=0.2$ in (a-b), and $t_1=-0.2$ in (c-d). The other parameters are chosen as $L = 50, m=40, n=20, g=1, t_2 = 1,$ and $\gamma = 0.5$.}
\label{f3}
\end{figure}

\begin{figure}[tbp]
\centering
\includegraphics[width=8.8cm]{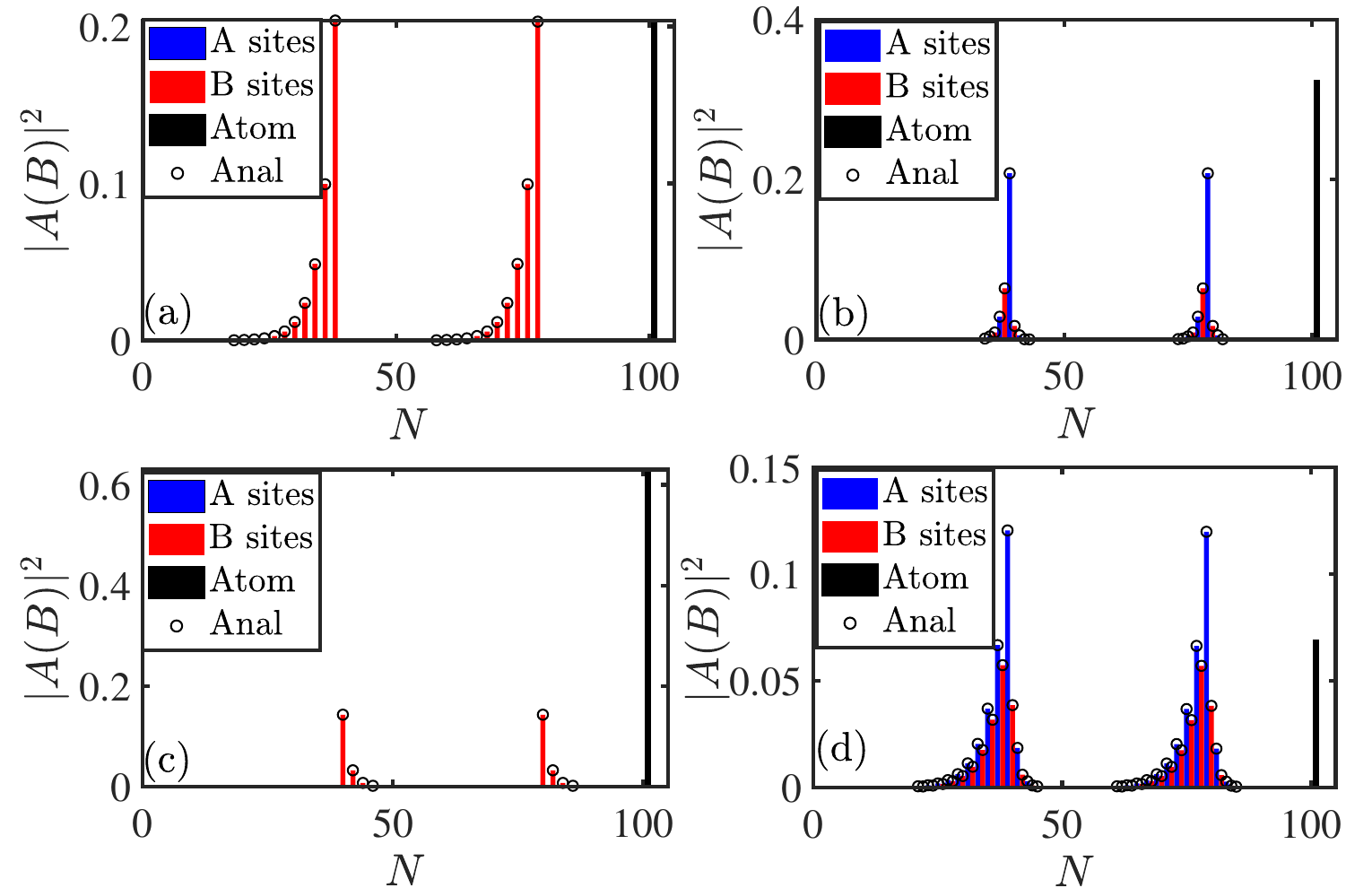}
\caption{The probability distributions of the system ~\eqref{3b} in bound states with $A-A$ coupling as a function of lattice site $N$. (a) and (c) are plotted for the probability distribution of zero mode, and corresponding wave functions do not distribute on $A$ sites. (b) and (d) are plotted for the probability distribution of the upper bound state.  The system in the bulk topological non-trivial phase with $t_1=0.2$ in (a-b) and the bulk topological trivial phase with $t_1=1.6$ in (c-d). The other parameters are set as $L = 50, m=40, n=20, g=1, t_2 = 1,$ and $\gamma = 0.5$.}
\label{f4}
\end{figure}

To begin with, we define probability distributions as modular square of the probability amplitude $|A(B)|^2$ in real space. The probability distributions for zero modes and  upper bound states as a function of lattice site $N$ with different parameters are shown in Fig.~\ref{f3} for $A-B$ coupling and in Fig.~\ref{f4} for $A-A$ coupling. The probability distribution on the giant atom is set at $N=2L+1=101$. The bars represent the numerical results and the empty circles represent the analytical results. Simple algebra shows that the analytical results of the probability amplitudes of bound states with energy $E$ take (see Appendix~\ref{A2} for details)
\begin{subequations}
\begin{equation}
\frac{A_{l} }{ U_{e}} =\frac{(-1)^{y+1}\left(T\tau_1^{|l-n|}+g\tau_2^{|l-m-1|}+Y_2\tau_2^{|l-m|}\right) }{\sqrt{x^{2}-4(t_1+\gamma)(t_2-\gamma)}}, \\
\label{8a}
\end{equation}
\begin{equation}
\frac{B_{l} }{ U_{e}}  =\frac{(-1)^{y+1}\left({T\tau_2^{|l-m|}+g\tau_1^{|l-n+1|}+Y_{1}\tau_1^{|l-n|}}\right)}{\sqrt{x^{2}-4(t_1+\gamma)(t_1-\gamma)}}
\label{8b}
\end{equation}
\end{subequations}
for the $A-B$ coupling,  and
\begin{subequations}
\begin{equation}
\frac{A_{l} }{ U_{e}}  =\frac{(-1)^{y+1}T}{\sqrt{x^{2}-4(t_1+\gamma)(t_1-\gamma)}}\left(\tau_1^{|l-n|}+\tau_3^{|l-m|}\right), \\
\label{9a}
\end{equation}
\begin{equation}
\begin{aligned}
\frac{B_{l} }{ U_{e}}  =&\frac{(-1)^{y+1}}{\sqrt{x^{2}-4(t_1+\gamma)(t_1-\gamma)}}\left\{Y_{1}\left(\tau_1^{|l-n|}+\tau_3^{|l-m|}\right)\right.\\
&\left.+g\left(\tau_1^{|l-n+1|}+\tau_3^{|l-m+1|}\right)\right\}
\label{9b}
\end{aligned}
\end{equation}
\end{subequations}
for the $A-A$ coupling. Here, $y = \theta(x)$ is the step function, and $x=(E^2-(t_1+\gamma)(t_1-\gamma)-t_2^2)/t_2$. We define $T=gE/t_2$, $Y_{1(2)}=g(t_1\mp\gamma)/t_2$,
$\tau_1=a\,(b)$ for $l\geq n\,(l<n)$, $\tau_2=a\,(b)$ for $l>m$\,($l\leq m$), $\tau_3=a\,(b)$ for $l\geq m$\,($l<m$), and $a=(x\pm\sqrt{x^2-4(t_1+\gamma)(t_1-\gamma)})/2(t_1+\gamma), b=(x\pm\sqrt{x^2-4(t_1+\gamma)(t_1-\gamma)})/2(t_1-\gamma)$ with $+(-)$ corresponding to $x<-2|t_1|$ ($x>2|t_1|$).

For the system with $A-B$ coupling, the zero mode present in the bulk topological non-trivial phase of nonreciprocal SSH model with $-t_2+\gamma<t_1<t_2-\gamma$. Setting  $E=0$, the probability distribution of the zero mode can be obtain from Eqs.~\eqref{8a} and~\eqref{8b},
\begin{equation}
\begin{aligned}
&A_{l} / U_{e}=Y_{3}\times \begin{cases}\left(-\frac{t_{1}-\gamma }{ t_{2}}\right)^{(l-m)}, & (l>m), \\
0, & (l \leq m),\end{cases} \\
&B_{l} / U_{e}=Y_{4}\times \begin{cases}\left(-\frac{t_{1}+\gamma }{ t_{2}}\right)^{(n-l)}, & (l<n), \\
0, & (l \geq n),\end{cases}
\label{10}
\end{aligned}
\end{equation}
where $Y_3=g/(t_1-\gamma)$, and $Y_4=g/(t_1+\gamma)$. We find that the probability amplitudes in the wave function   satisfy the asymmetrical distribution $|A_l|\neq|B_{m+n-l}|$. Assuming  $\gamma>0$, we have $|(t_1-\gamma)/t_2|<|(t_1+\gamma)/t_2|$ for $t_1>0$, and $|(t_1+\gamma)/t_2|<|(t_1-\gamma)/t_2|$ for $t_1<0$, which suggests that the system  mainly occupies  $B$ sites on the left side of the giant atom when $t_1>0$, and occupies the $A$ sites on the right side of the giant atom when $t_1<0$. Obviously, the spatial distributions of wave functions on the left and right sides of the giant atom are the same for the critical point with $t_1=0$. Besides, as the increase of $g$, the probability of the system occupying   the giant atom are gradually suppressed, but the spatial symmetry of zero modes on the SSH chain can not be changed.

In Figs.~\ref{f3}(a) and~\ref{f3}(c), we show the probability distributions of zero modes for $A-B$ coupling as a function of lattice site $N$ with $t_1=0.2$ and $t_1=-0.2$, respectively. In this case, the zero modes mainly locate on the left (right) sides of the giant atom with $t_1=0.2$ ($t_1=-0.2$), and satisfy the chiral symmetry~\cite{Nori2021}. It is clear that the analytical results given by Eq.~\eqref{10} are in good agreement  with the numerical results. In Figs.~\ref{f3}(b) and~\ref{f3}(d), we show the corresponding distributions of the upper bound state with $t_1=0.2$ and $t_1=-0.2$, respectively. Although the amplitude distributions of wave function on sites are lack of spatial symmetry, the non-reciprocity can induce a supremacy  that makes the upper bound states to distribute on the left (right) of coupling point more than the right (left) of coupling point for $t_1=0.2$ ( $t_1=-0.2$). Similarly, the analytical results given by Eq.~\eqref{8a} and Eq.~\eqref{8b} are in good agreement  with the numerical results.

For the system with $A-A$ coupling, zero modes are always there regardless of the value of $t_1$. According to Eq.~\eqref{9a} and Eq.~\eqref{9b} , we can get the probability distribution of the zero mode, which gives rise to $A_{l} =0$, and
\begin{equation}
B_{l} / U_{e}=Y_{4} \times \begin{cases}\left(\frac{-t_{1}-\gamma }{ t_{2}}\right)^{(n-l)}+\left(\frac{-t_{1}-\gamma }{t_{2}}\right)^{(m-l)},~~~~(l<n),
\\ \left(\frac{-t_{1}-\gamma }{ t_{2}}\right)^{(m-l)},~~~~~~~~~~~~~~~~~~~~~~~~(n \leq l<m), \\
0,~~~~~~~~~~~~~~~~~~~~~~~~~~~~~~~~~~~~~~~~~~~~~~~~~~~~~~~(m \leq l),\end{cases}
\label{11}
\end{equation}
in the bulk topological non-trivial phase of the SSH model with $-t_2+\gamma<t_1<t_2-\gamma$ and
\begin{equation}
B_{l} / U_{e}=-Y_{4} \times \begin{cases}0,~~~~~~~~~~~~~~~~~~~~~~~~~~~~~~~~~~~~~~~~~~~~~~~~~~~~~(l<n),\\
\left(\frac{-t_{2}} { t_{1}+\gamma}\right)^{l-n},~~~~~~~~~~~~~~~~~~~~~~~~~~~(n \leq l<m), \\
\left(\frac{-t_{2} }{ t_{1}+\gamma}\right)^{l-n}+\left(\frac{-t_{2} }{ t_{1}+\gamma}\right)^{l-m},~~~~~~~~~~(m \leq l),\end{cases}
\label{12}
\end{equation}
in the bulk topological trivial phase of the SSH model with $t_1>t_2+\gamma$ or $t_1<-t_2-\gamma$. As for $-t_2-\gamma<t_1<-t_2+\gamma$ and $-t_2-\gamma<t_1<-t_2+\gamma$, the analytical results of zero model are not given. Note that the amplitudes of the wave functions only distribute on $B$ sites on the left sides of the giant atom or between two coupling points with $-t_2+\gamma<t_1<t_2-\gamma$, and distribute on $B$ sites on the right sides of the giant atom or between two coupling points with $t_1>t_2+\gamma$ or $t_1<-t_2-\gamma$. Similarly, distributions of wave function can also be changed by tuning the parameters of the system. Taking  $t_1$ $t_2$ and $\gamma$ to be positive  leads to $\gamma/t_2<|(t_1+\gamma)/t_2|<1$ in the bulk topological non-trivial phase, and $0<|t_2/(t_1+\gamma)|<t_2/(t_2+2\gamma)$ in the bulk topological trivial phase. Taking $t_2\leq2\gamma$  yields $\gamma/t_2\geq t_2/(t_2+2\gamma)$. This indicates that the decay of distributions in the bulk topological non-trivial phase are always slower than the bulk topological trivial phase.

In order to verify the above analysis, in Figs.~\ref{f4}(a) and~\ref{f4}(c), we show  the probability distributions of zero modes for $A-A$ coupling as a function of lattice site $N$ with $t_1=0.2$ and $t_1=1.6$, respectively. The results in Figs.~\ref{f4}(a) and~\ref{f4} (c) are obtained by numerical simulations and they show that $|B_l|\approx|B_{l+m-n}|$ are satisfied. The decay of distributions with $t_1=0.2$ are slower than the distributions with $t_1=1.6$. In Figs.~\ref{f3}(b) and~\ref{f3}(d), we show the corresponding distributions of the upper bound state with $t_1=0.2$ and $t_1=1.6$, respectively. It is easy to find  that they satisfy $|A(B)_l|\approx|A(B)_{l+m-n}|$, and the non-reciprocity makes upper bound states to distribute on the left of coupling point more than the right of coupling point.

\begin{figure}[tbp]
\centering
\includegraphics[width=8.8cm]{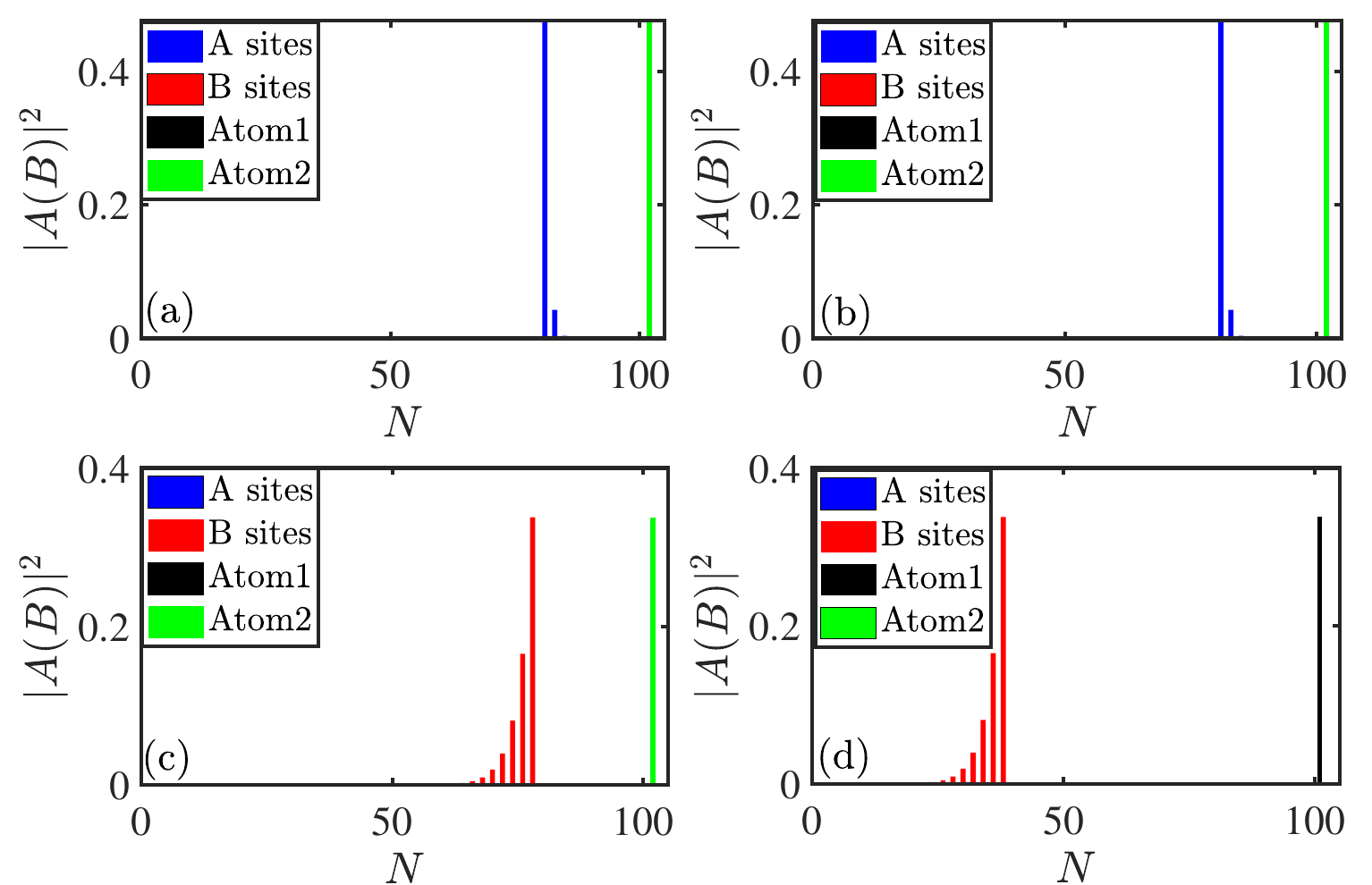}
\caption{The probability distributions of zero modes for the system that two small atoms couple to a nonreciprocal SSH chain as a function of lattice site $N$. (a) and (b) are plotted for the $A-B$ coupling. (c) and (d) are plotted for the $A-A$ coupling. Here, $L = 50, m=40, n=20, g=1, t_1=0.2, t_2 = 1,$ and $\gamma = 0.5$.}
\label{f5}
\end{figure}

In Fig.~\ref{f3} and Fig.~\ref{f4}, we present rich and intriguing zero modes for the system composed of a giant atom and a nonreciprocal SSH chain. Note that zero modes localize at two effective boundaries. While for a nonreciprocal SSH chain with OBC, zero modes only localize at one of the two boundaries~\cite{Yaoa2018}. These two cases are different, because the giant atom not only induces two effective boundaries, but also couples them. In order to show the validity of the above conclusions, we plot the probability distributions of zero modes for the system that two small atoms couple to a nonreciprocal SSH chain. The coupling points locate at sites $m$ and $n$, respectively. As shown in Fig.~\ref{f5}, there are two zero modes for small-atom case. Both modes localize at the right coupling point for $A-A$ coupling, and one for each coupling point for $A-B$ coupling with $m=40, n=20, g=1, t_1=0.2, t_2 = 1,$ and $\gamma = 0.5$. It is clearly shown that although two small atoms as defects induce two effective boundaries, zero modes localize at only one of the two boundaries. Note that this case is similarly to the zero mode with OBC. Hence, it can be seen that the zero modes for systems with giant atom are very unique, because nonlocal couplings weaken the localization.

\section{localization of eigenstates}\label{sec4}
\begin{figure}[tbp]
\centering
\includegraphics[width=9cm]{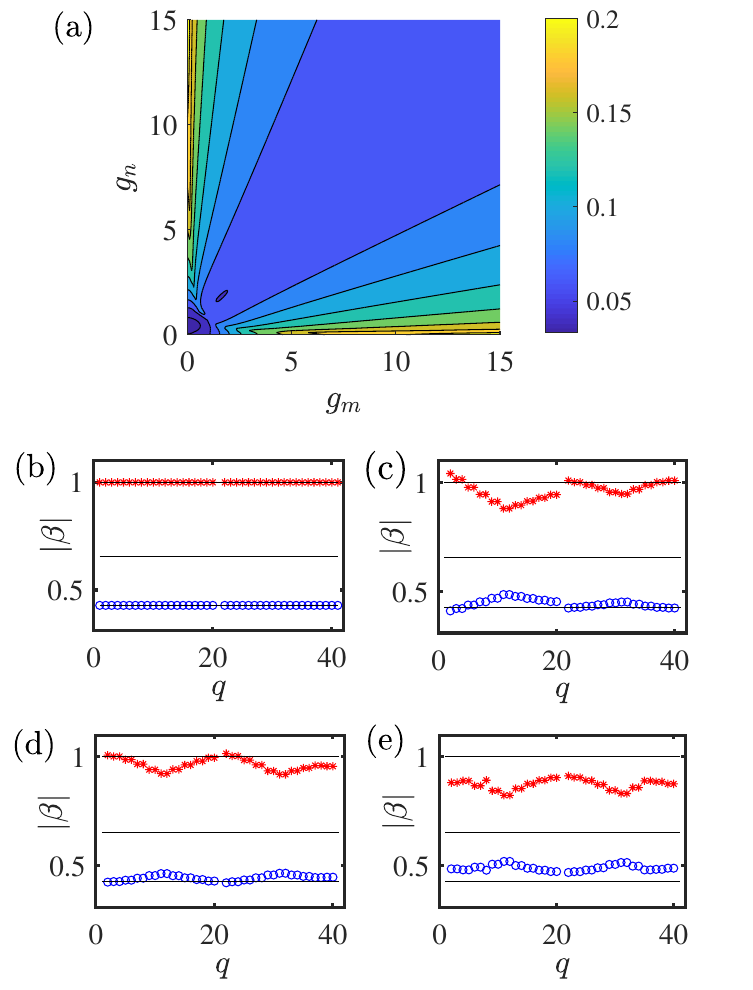}
\caption{(a) $\overline{\mathrm{IPR}}$ on the parameter space of $g_m$ and $g_n$ for the system~\eqref{3a} with $A-B$ coupling given by Eq.~\eqref{13}. (b-e) $|\beta_1|$ (red star) and $|\beta_2|$ (blue circle) as a function of $q$ given by Eq.~\eqref{15} with $(g_m, g_n)=(0,0), (1,1), (13,13),$ and $(13,1)$, respectively. The three reference lines from top to bottom correspond to $|\beta|=1, \sqrt{|(t_1-\gamma)/(t_1+\gamma)|}$, and $|(t_1-\gamma)/(t_1+\gamma)|$. The parameters shared by all of the figures are $L = 20, m=n=10, t_1=0.2, t_2 = 1,$ and $\gamma = 0.5$.}
\label{f6}
\end{figure}

Our aim is to study the localization of eigenstates, and we only interest in  the wave functions living at the SSH chain, so distributions on the giant atom are not considered. To this aim, we calculate the modified averaged inverse participation ratio $\overline{\mathrm{IPR}}$,
\begin{equation}
\overline{\mathrm{IPR}}=\frac{1}{N}\sum_{q=1}^{N}\mathrm{IPR}_q
\label{13}
\end{equation}
with
$\mathrm{IPR}_q=\sum_{N=1}^{2L}|\psi^q_{N}|^4/(\sum_{N=1}^{2L}|\psi^q_{N}|^2)^2$ and $\psi^q_{N}$ is the probability amplitude at site $N$ of the $q$th right eigenstate of Hamiltonian in real space. $\overline{\mathrm{IPR}}$ will enhance as the increase of the localization. In Fig.~\ref{f6}\,(a), we show $\overline{\mathrm{IPR}}$ on the parameter space of $g_m$ and $g_n$ for the giant-atom case with $A-B$ coupling ($A-A$ coupling is similar). $g_m$ and $g_n$ are the atom-chain coupling strength, where $m$ and $n$  denote the coupling sites. Note that $\overline{\mathrm{IPR}}$ is suppressed when $g_m=g_n$, and can be lifted as the increase of $\delta g=|g_n-g_m|$, which indicates that the localization will be suppressed for the same coupling strength setting. Particularly, for the case of $g_n\neq0~(g_m=0)$ or $g_m\neq0~(g_n=0)$, i.e., like a small-atom case, $\overline{\mathrm{IPR}}$ will significant enhance as the increase of $g_n$ or $g_m$, which suggests that nonlocal couplings weaken localization of eigenstates for giant-atom case.

In the last section, we have shown the behavior of the bound states. Next, to analyze the localization of bulk states in detail, let us consider the real-space eigen-equation of the system. In the bulk of chain, it satisfies
\begin{equation}
\begin{gathered}
t_2\psi_{B,l-1}+[t_1+\gamma]\psi_{B,l}=E\psi_{A,l},\\
[t_1-\gamma]\psi_{A,l}+t_2\psi_{A,l+1}=E\psi_{B,l}.
\label{14}
\end{gathered}
\end{equation}
With $(\psi_{A,l}, \psi_{B,l})=\beta^l(\psi_A, \psi_B)$, it yields
\begin{equation}
\beta_{1,2}(E)=\frac{[\Delta\pm\sqrt{\Delta^2-4t_2^2(t_1^2-\gamma^2)}]}{2t_2(t_1+\gamma)},
\label{15}
\end{equation}
where $\Delta=E^2+\gamma^2-t_1^2-t_2^2$, and $+(-)$ corresponds to $\beta_{1}\left(\beta_{2}\right)$. In Fig.~\ref{f6}\,(b-e), we show $|\beta_1|$ (red star) and $|\beta_2|$ (blue circle) as a function of $q$ with different coupling strengths $(g_m, g_n)=(0,0), (1,1), (13,13),$ and $(13,1)$. For the system under OBC without the giant atom, $|\beta_{1,2}|=\sqrt{|(t_1-\gamma)/(t_1+\gamma)|}$~\cite{Yaoa2018} corresponds to the line in the middle, and the wave functions take the form of skin states. While for the system under PBC without the giant atom, $|\beta_{1}|=1$, $|\beta_2|=|(t_1-\gamma)/(t_1+\gamma)|$ corresponds to the line in the above and below, respectively, and the wave functions take the form of skin-free states.  When coupling strengths $g_m=g_n=0$, the bulk states obviously take the form of skin-free states (Fig.~\ref{f6} (b)). Counter-intuitively, $|\beta_{1,2}|$ cross the line in the above and below, respectively as $g_m=g_n=1 ( or~ 13)$ (Fig.~\ref{f6} (c-d)), which suggests that bulk states might localize at the left ($|\beta_1|>1$) or the right ($|\beta_1|<1$) chain-atom coupling sites as long as coupling strengths are the same. Note that eigenstates are in a relatively weak localization regime in this case (Fig.~\ref{f6} (a)). This bipolar localization inevitably leads to Bloch-like states albeit broken translational invariance, but the bipolar localization will disappear in a relatively strong localization regime $(g_m, g_n)=(13,1)$ as shown in Fig.~\ref{f6} (e).
\begin{figure}[tbp]
\centering
\includegraphics[width=9cm]{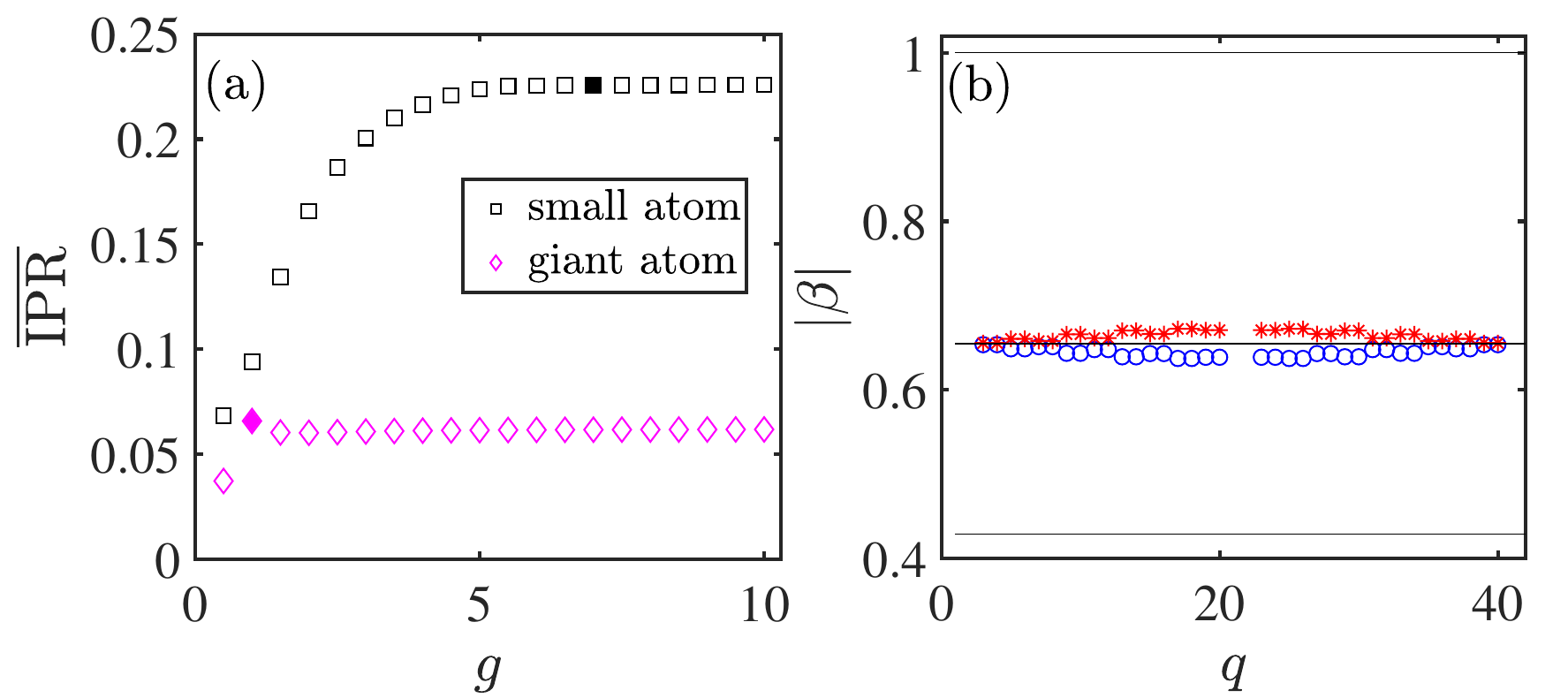}
\caption{(a) $\overline{\mathrm{IPR}}$ as a function of $g$ given by~Eq.~\eqref{13} for the system that a giant atom (magenta rhombus) or two small atoms (black square) coupls to a nonreciprocal SSH chain. (b) $|\beta_1|$ (red star) and $|\beta_2|$ (blue circle) as a function of $q$ given by Eq.~\eqref{15} with $g=7$ corresponds to black filled square in (a). The parameters are chosen as $L = 20, m=n=10, t_1=0.2, t_2 = 1,$ and $\gamma = 0.5$.}
\label{f7}
\end{figure}

In order to further illustrate that nonlocal couplings have significant effect on the localization of eigenstates, we show $\overline{\mathrm{IPR}}$ as a function of the coupling strengths $g_m=g_n=g$ for the system that a giant atom or two small atoms couples to a nonreciprocal SSH chain in Fig.~\ref{f7}\,(a). Note that $\overline{\mathrm{IPR}}$ for giant-atom case reaches peak at $g=1$ corresponding to magenta filled rhombus in Fig.~\ref{f7}\,(a), and even higher than the case with a bigger $g$. Comparing to small-atom case, it clear that nonlocal coupling weakens localization of eigenstates.

In Fig.~\ref{f7}\,(b), we show $|\beta_1|$ (red star) and $|\beta_2|$ (blue circle) as a function of $q$ with $g=7$ for small-atom case, and corresponding to black filled square in Fig.~\ref{f7}\,(a). It can be seen that $|\beta_{1,2}|\approx\sqrt{|(t_1-\gamma)/(t_1+\gamma)|}$ are similar to the case with OBC, which implies that bulk states are like the form of skin states. On the contrary, it can not be realized for giant-atom case even for a bigger $g=13$ as shown in Fig.~\ref{f6}\,(d).

\section{lyapunov exponent}\label{sec5}
\begin{figure}[tbp]
\centering
\includegraphics[width=9cm]{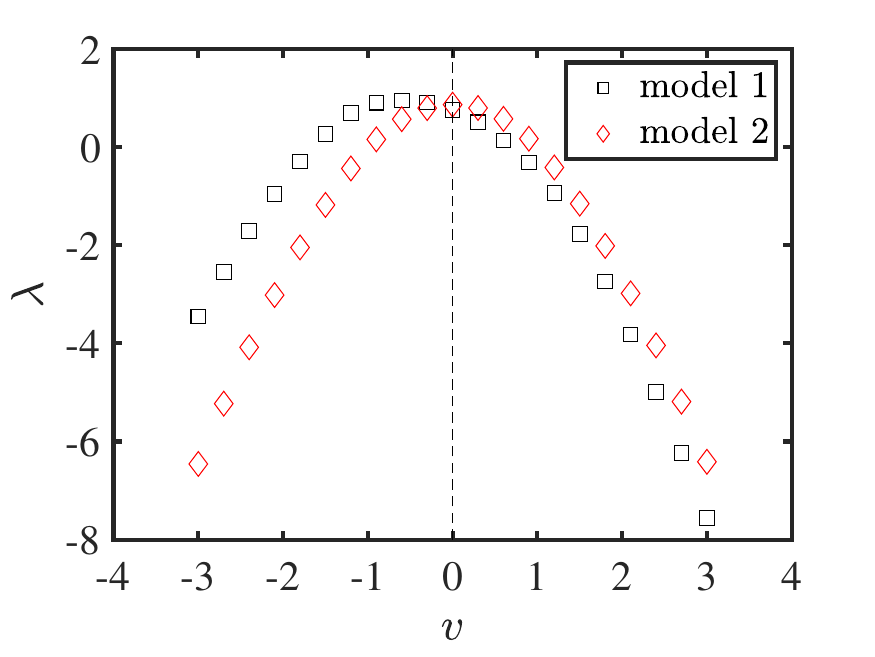}
\caption{Lyapunov exponent $\lambda(v)$ as a function of $v$ given by~Eq.~\eqref{16} for the system~\eqref{3a} (black squares) or the system~\eqref{17} (red diamonds) with $A-B$ coupling, respectively. The parameters are set as $L = 401, m=n=1, t_1=0.6, t_2 = 1, \gamma = 1$, and $\delta=1$.}
\label{f8}
\end{figure}
In order to predict the localization of bulk states, we calculate the Lyapunov exponent $\lambda(v)$ in the long-time dynamics in real space far from edges,
\begin{equation}
\lambda(v)=\lim _{t \rightarrow \infty} \frac{\log |\psi(t)|}{t},
\label{16}
\end{equation}
where the amplitude $|\psi(t)\rangle=A_{(L+1)/2+n}(t)$ [or similarly$ |\psi(t)\rangle=B_{(L+1)/2+n}(t) ]$ varies along the space-time path $n=vt$, and $v$ is the drift velocity~\cite{Dee1983,Longhi2013,Longhi2019}. Recent research shows that if $\lambda(v)$ does not reach its largest value at zero drift velocity, then the non-Hermitian system exhibits the NHSE~\cite{Longhi2019}. Clearly, Lyapunov exponent can act as a witness of defect-induced localized bulk states in the same way. For example, the existence of the giant atom induced localized bulk states can be captured by the Lyapunov exponent exhibiting its largest value at a nonvanishing drift velocity.

The Lyapunov exponent in the long-time dynamics is numerically computed by solving the time-dependent Schr\"odinger equation. In Fig.~\ref{f8}, model $1$ with localized bulk states corresponds to the nonreciprocal system~\eqref{3a} with $A-B$ coupling ($A-A$ coupling is similar). As a comparison, model $2$ without localized bulk states is described by
\begin{equation}
H^{'}_{AB}=H^{'}_{\mathrm{SSH}}+H_{I,AB},
\label{17}
\end{equation}
with
\begin{widetext}
\begin{equation}
\begin{aligned}
H^{'}_{\mathrm{SSH}}=&\sum_{l=1}^L[t_1\hat C^{\dag}_{A,l}\hat C{_{B,l}}+t_1\hat C{^{\dag}_{B,l}}\hat C{_{A,l}}
+t_2\hat C{^{\dag}_{A,l+1}}\hat C{_{B,l}}
+t_2\hat C{^{\dag}_{B,l}}\hat C{_{A,l+1}}+i\delta \hat C{^{\dag}_{A,l}}\hat C{_{A,l}}-i\delta \hat C{^{\dag}_{B,l}}\hat C{_{B,l}}],
\label{18}
\end{aligned}
\end{equation}
\end{widetext}
where gain or loss ($\pm i\delta$) of the sublattice $A(B)$ leads to  non-Hermiticity. The initial state is prepared in the bulk of the lattice, i.e., $A_{(L+1)/2+n}(0)=B_{(L+1)/2+n}(0)=\delta_{n,0}$, and we chose  a long lattice ($L=401$) to avoid the wave packet reaching the effective boundary at the observation time $t=50$. This time is sufficient to compute $\lambda(v)$ with good accuracy. It is clearly shown that Lyapunov exponent exhibiting its largest value at a nonvanishing drift velocity can act as a witness of the existence of localized bulk states.

\section{summary}\label{sec6}
In summary, we have studied a giant atom coupled to two points of a nonreciprocal SSH chain. We show the spectrum structures of the system, and give a condition for the emergence of the zero mode. The interplay of nonreciprocal hopping and the nonlocal couplings can induce asymmetric zero modes. It is clear that the analytical results of zero modes are precise and almost identify with the numerical results. The features of zero modes are unique and obviously different from the case with OBC or two small atoms. Counter-intuitively, we uncover that bulk states might localize at the left or the right chain-atom coupling sites in weak localization regimes, and the localization is obviously weaker than the case with small-atom or OBC even in strong coupling regimes. The above results suggest that nonlocal coupling of giant atoms to a nonreciprocal SSH chain weakens localization of both zero modes and bulk states. We also show that Lyapunov exponent exhibiting its largest value at a nonvanishing drift velocity can act as a witness of the localized bulk states.

\section{acknowledgments}
The authors acknowledge Weijun Cheng for helpful comments. This work was supported by the National Natural Science Foundation of China (NSFC) under Grants No. 12175033, No. 12147206 and National Key R$\&$D Program of China (No. 2021YFE0193500).

\appendix
\addcontentsline{toc}{section}{Appendices}\markboth{APPENDICES}{}
\begin{subappendices}
\section{energy equation}
\label{A1}
Consider a system consisting of a giant atom coupled to a nonreciprocal SSH chain via A-B coupling, the Hamiltonian of the system in the momentum space reads,
\begin{equation}
\begin{aligned}
H_{AB}(k)=&H_{\mathrm{SSH}}(k)+H_{I, A B}(k)\\
=&\sum_{k}[[(t_1+\gamma)+t_2e^{-ik}]\hat C_{A,k}^\dag\hat C_{B,k}
+[(t_1-\gamma)\\
&+t_2e^{ik}]\hat C_{B,k}^\dag\hat C_{A,k}]
+\frac{g}{\sqrt{L}} \sum_{k}[|e\rangle\langle g|(\hat{C}_{A, k} e^{i k n}\\
&+\hat{C}_{B, k} e^{i k m})+\text { H.c. }].
\label{a1}
\end{aligned}
\end{equation}
The time-independent Sch\"{o}dinger equation $H_{AB}(k)|\psi\rangle = E|\psi\rangle$ together with Eq.~\eqref{4} leads to,
\begin{equation}
\begin{aligned}
&E U_{e}=\frac{g}{\sqrt{L}} \sum_{k}\left(A_{k} e^{i k n}+B_{k} e^{i k m}\right), \\
&E A_{k}=\left(t_{1}+\gamma+t_{2} e^{-i k}\right) B_{k}+\frac{g}{\sqrt{L}} e^{-i k n} U_{e}, \\
&E B_{k}=\left(t_{1}-\gamma+t_{2} e^{i k}\right) A_{k}+\frac{g}{\sqrt{L}} e^{-i k m} U_{e}.
\label{a2}
\end{aligned}
\end{equation}
Eliminating $A_k$,$B_k$ and $U_e$ in  Eq.~\eqref{a2}, we obtain the equation for $E$,
\begin{equation}
\begin{aligned}
E &=\frac{2 g^{2}}{L} \sum_{k} (E+t_{1} \cos [k(m-n)]-\gamma \sin [k(m-n)]\\
&+t_{2} \cos [k(m-n+1)])/(E^{2}-\omega_{k}^{2}) \\
&=\frac{g^{2}}{\pi} \int_{-\pi}^{\pi} d k (E+t_{1} \cos [k(m-n)]-\gamma \sin [k(m-n)]\\
&+t_{2} \cos [k(m-n+1)])/(E^{2}-\omega_{k}^{2}).
\label{a3}
\end{aligned}
\end{equation}
To obtain  the zero-mode, we set $E = 0$ and write  $z_1 = e^{ik}$ and $z_2 = e^{-ik}$, the right-hand side of the last equation can be simplified via Residue theorem (although $E=0$, we keep it in the equation for clarity of expression)
\begin{equation}
\begin{aligned}
&\frac{g^{2}}{\pi} \int_{-\pi}^{\pi} d k (E+t_{1} \cos [k(m-n)]-\gamma \sin [k(m-n)]\\
&+t_{2} \cos [k(m-n+1)])/(E^{2}-\omega_{k}^{2})\\
=&\frac{-g^2}{2\pi i}[\oint_{|z_1|=1}dz_1\frac{(t_1-\gamma)z_1^{m-n}+t_2 z_1^{m-n+1}}{(t_1+\gamma)t_2(z_1+\frac{t_2}{t_1+\gamma})(z_1+\frac{t_1-\gamma}{t_2})}\\
&+\oint_{|z_2|=1}dz_2\frac{(t_1+\gamma)z_2^{m-n}+t_2 z_2^{m-n+1}}{(t_1-\gamma)t_2(z_2+\frac{t_2}{t_1-\gamma})(z_2+\frac{t_1+\gamma}{t_2})}]\\
=& \begin{cases}
\frac{-g^2}{t_1-\gamma}\left(\frac{-t_{2}}{t_{1}-\gamma}\right)^{m-n}, & (-t_2-\gamma<t_{1}< -t_{2}+\gamma), \\\\
0, & \left(-t_{2}+\gamma < t_{1}<t_2-\gamma\right),\\\\
\frac{-g^2}{t_1+\gamma}\left(\frac{-t_{2}}{t_{1}+\gamma}\right)^{m-n}, & (t_2-\gamma<t_{1}< t_{2}+\gamma), \\\\
\frac{-g^2}{t_1-\gamma}\left(\frac{-t_{2}}{t_{1}-\gamma}\right)^{m-n},  &(t_{1}<-t_2-\gamma,\\
+\frac{-g^2}{t_1+\gamma}\left(\frac{-t_{2}}{t_{1}+\gamma}\right)^{m-n},   &or~t_1>t_{2}+\gamma). \\
\end{cases}
\label{a4}
\end{aligned}
\end{equation}
Clearly, when $-t_2+\gamma<t_1<t_2-\gamma$, we can find the zero mode $E=0$.

Similarly, for the system with $A-A$ coupling, the interaction Hamiltonian in the momentum space reads
\begin{equation}
\begin{aligned}
{H_{I,AA}}(k) = \frac{{g}}{{\sqrt L }}\sum\limits_k { {|e\rangle \langle g|{{\hat C}_{A,k}}\left( {{e^{ikn}} + {e^{ikm}}} \right) + {\rm{ H}}{\rm{.c}}{\rm{. }}} },
\label{a5}
\end{aligned}
\end{equation}
and $H_{AA}(k)|\psi\rangle = E|\psi\rangle$ leads to
\begin{equation}
\begin{aligned}
&E U_{e}=\frac{g}{\sqrt{L}} \sum_{k} A_{k}\left(e^{i k n}+e^{i k m}\right), \\
&E A_{k}=\left(t_{1}+\gamma+t_{2} e^{-i k}\right) B_{k}+\frac{g}{\sqrt{L}}\left(e^{-i k n}+e^{-i k m}\right) U_{e}, \\
&E B_{k}=\left(t_{1}-\gamma+t_{2} e^{i k}\right) A_{k}.
\label{a6}
\end{aligned}
\end{equation}
Some algebras shows that the energy $E$ satisfies
\begin{equation}
E=\frac{2 g^{2}}{L} \sum_{k}\left(\frac{E(1+\cos [k(m-n)])}{E^{2}-\omega_{k}^{2}}\right).
\label{a7}
\end{equation}
Evidently, $E = 0$ is always the solution of the Eq.~\eqref{a7}.

\section{states outside the band}
\label{A2}
For the system with $A-B$ couplings, in accordance  to the Eq.~\eqref{a2}, the probability amplitudes $A_k$ and $B_k$  satisfy,
\begin{equation}
\begin{aligned}
A_{k} / U_{e} &=\frac{g}{\sqrt{L} t_{2}}\left(E e^{-i k n}+\left(t_{1}+\gamma+t_{2} e^{-i k}\right) e^{-i k m}\right) f(k), \\
B_{k} / U_{e} &=\frac{g}{\sqrt{L} t_{2}}\left(E e^{-i k m}+\left(t_{1}-\delta+t_{2} e^{i k}\right) e^{-i k n}\right) f(k), \\
\label{b1}
\end{aligned}
\end{equation}
where $f(k)=1/(x-(t_1+\gamma)e^{ik}-(t_1-\gamma)e^{-ik})$ and $x=(E^2-(t_1+\gamma)(t_1-\gamma)-t_2^2)/t_2$. By Fourier expansion for $f(k)$,
\begin{equation}
f(k)=a_{0}+\sum_{p}\left[\left(\frac{a_{p}-i b_{p}}{2}\right) e^{i k p}+\left(\frac{a_{p}+i b_{p}}{2}\right) e^{-i k p}\right],
\label{b2}
\end{equation}
with
\begin{equation}
\begin{aligned}
a_{0} &=\frac{1}{2\pi} \int_{-\pi}^{\pi} f(k) d k, \\
\frac{a_{p}+i b_{p}}{2} &=\frac{1}{2 \pi} \int_{-\pi}^{\pi} f(k) e^{i k p} d k ,\\
\frac{a_{p}-i b_{p}}{2} &=\frac{1}{2 \pi} \int_{-\pi}^{\pi} f(k) e^{-i k p} d k,
\label{b3}
\end{aligned}
\end{equation}
we derive  $f(k)$ as follows
\begin{equation}
f(k)=\frac{(-1)^{y+1}}{\sqrt{x^{2}-4(t_1+\gamma)(t_1-\gamma)}}[1+\sum_{p=1}^{L}\left(e^{-i k p} a^{p}+e^{i k p} b^{p}\right)],
\label{b4}
\end{equation}
where $y = \theta(x)$ is the step function, and $a=(x+\sqrt{x^2-4(t_1+\gamma)(t_1-\gamma)})/2(t_1+\gamma), b=(x+\sqrt{x^2-4(t_1+\gamma)(t_1-\gamma)})/2(t_1-\gamma)$ for $x<-2|t_1|$ or $a=(x-\sqrt{x^2-4(t_1+\gamma)(t_1-\gamma)})/2(t_1+\gamma), b=(x-\sqrt{x^2-4(t_1+\gamma)(t_1-\gamma)})/2(t_1-\gamma)$ for $x>2|t_1|$.
Substituting the above results into Eq.~\eqref{b1},  we will get the probability amplitude in real space by inverse Fourier transformation. The result is,

\begin{equation}
\begin{aligned}
&A_{l} / U_{e} =\frac{1}{\sqrt{L}} \sum_{k} e^{i k l} A_{k} / U_{e} \\
&=\frac{(-1)^{y+1}\left(T\tau_1^{|l-n|}+g\tau_2^{|l-m-1|}+Y_2\tau_2^{|l-m|}\right) }{\sqrt{x^{2}-4(t_1+\gamma)(t_2-\gamma)}}, \\
\label{b5}
\end{aligned}
\end{equation}

\begin{equation}
\begin{aligned}
&B_{l} / U_{e} =\frac{1}{\sqrt{L}} \sum_{k} e^{i k l} B_{k} / U_{e} \\
&=\frac{(-1)^{y+1}\left({T\tau_2^{|l-m|}+g\tau_1^{|l-n+1|}+Y_{1}\tau_1^{|l-n|}}\right)}{\sqrt{x^{2}-4(t_1+\gamma)(t_1-\gamma)}},
\label{b6}
\end{aligned}
\end{equation}
where $T=gE/t_2$, $Y_1=g(t_1-\gamma)/t_2$, $Y_2=g(t_1+\gamma)/t_2$. $\tau_1=a$ for $l\geq n$, $\tau_1=b$ for $l<n$. $\tau_2=a$ for $l>m$, $\tau_2=b$ for $l\leq m$.  For the zero modes  $E = 0$, we would have $x < -2|t_1|$, thus $a$ and $b$ are reduced to $-(t_1-\gamma)/t_2$ and $-(t_1+\gamma)/t_2$ for $-t_2+\gamma<t_1<t_2-\gamma$, respectively. $a$ and $b$ are reduced to $-t_2/(t_1+\gamma)$ and $-t_2/(t_1-\gamma)$ for $t_1>t_2+\gamma$ or $t_1<-t_2-\gamma$, respectively. Then  Eq.~\eqref{b5}and Eq.~\eqref{b6} can be simplified as,
\begin{equation}
\begin{aligned}
&A_{l} / U_{e}=Y_{3}\times \begin{cases}\left(-\frac{t_{1}-\gamma }{ t_{2}}\right)^{(l-m)}, & (l>m), \\
0, & (l \leq m),\end{cases} \\
\label{b7}
\end{aligned}
\end{equation}

\begin{equation}
\begin{aligned}
&B_{l} / U_{e}=Y_{4}\times \begin{cases}\left(-\frac{t_{1}+\gamma }{ t_{2}}\right)^{(n-l)}, & (l<n), \\
0, & (l \geq n),\end{cases}
\label{b8}
\end{aligned}
\end{equation}
where $Y_3=g/(t_1-\gamma)$,$Y_4=g/(t_1+\gamma)$.

Similarly, for the system with $A-A$ coupling,  Eq.~\eqref{a6} yields
\begin{equation}
\begin{aligned}
A_{k} / U_{e} &=\frac{g E}{\sqrt{L} t_{2}}\left(e^{-i k n}+e^{-i k m}\right) f(k), \\
B_{k} / U_{e} &=\frac{g\left(t_{1}-\gamma+t_{2} e^{i k}\right)}{\sqrt{L} t_{2}}\left(e^{-i k n}+e^{-i k m}\right) f(k).
\label{b9}
\end{aligned}
\end{equation}
By the inverse Fourier transformation, we can obtain
\begin{equation}
\begin{aligned}
A_{l} / U_{e} &=\frac{1}{\sqrt{L}} \sum_{k} e^{i k l} A_{k} / U_{e} \\
&=\frac{(-1)^{y+1} T}{\sqrt{x^{2}-4(t_1+\gamma)(t_1-\gamma)}}\left(\tau_1^{|l-n|}+\tau_3^{|l-m|}\right), \\
\label{b10}
\end{aligned}
\end{equation}
\begin{equation}
\begin{aligned}
B_{l} / U_{e} &=\frac{1}{\sqrt{L}} \sum_{k} e^{i k l} B_{k} / U_{e} \\
&=\frac{(-1)^{y+1}}{\sqrt{x^{2}-4(t_1+\gamma)(t_1-\gamma)}}\left\{Y_{1}\left(\tau_1^{|l-n|}+\tau_3^{|l-m|}\right)\right.\\
&\left.+g\left(\tau_1^{|l-n+1|}+\tau_3^{|l-m+1|}\right)\right\},
\label{b11}
\end{aligned}
\end{equation}
where  $\tau_3=a$ for $l\geq m$, $\tau_3=b$ for $l<m$. Then setting $E=0$, the Eq.~\eqref{b10} and Eq.~\eqref{b11} can be simplified as $A_{l} / U_{e}=0$,
\begin{equation}
B_{l} / U_{e}=-Y_{4} \times \begin{cases}0,~~~~~~~~~~~~~~~~~~~~~~~~~~~~~~~~~~~~~~~~~~~~~~~~~~~~~(l<n), \\
\left(\frac{-t_{2}} { t_{1}+\gamma}\right)^{l-n},~~~~~~~~~~~~~~~~~~~~~~~~~~~(n \leq l<m), \\
\left(\frac{-t_{2} }{ t_{1}+\gamma}\right)^{l-n}+\left(\frac{-t_{2} }{ t_{1}+\gamma}\right)^{l-m},~~~~~~~~~~(m \leq l),\end{cases}
\label{b12}
\end{equation}
for $t_1>t_2+\gamma$ or $t_1<-t_2-\gamma$, and
\begin{equation}
B_{l} / U_{e}=Y_{4} \times \begin{cases}\left(\frac{-t_{1}-\gamma }{ t_{2}}\right)^{(n-l)}+\left(\frac{-t_{1}-\gamma }{t_{2}}\right)^{(m-l)},~~~~(l<n),
\\ \left(\frac{-t_{1}-\gamma }{ t_{2}}\right)^{(m-l)},~~~~~~~~~~~~~~~~~~~~~~~~(n \leq l<m), \\
0,~~~~~~~~~~~~~~~~~~~~~~~~~~~~~~~~~~~~~~~~~~~~~~~~~~~~~~~(m \leq l),\end{cases}
\label{b13}
\end{equation}
for $-t_2+\gamma<t_1<t_2-\gamma$.
\end{subappendices}

\end{document}